\documentclass{emulateapj}
\usepackage{graphicx}
 
\slugcomment{Draft Version June 5, 2009}

\begin{document}


\title{
  DISCOVERY OF PULSATIONS FROM THE PULSAR J0205+6449 IN SNR 3C~58 \\
  WITH THE FERMI GAMMA-RAY SPACE TELESCOPE }

\author{
A.~A.~Abdo\altaffilmark{1,2}, 
M.~Ackermann\altaffilmark{3}, 
M.~Ajello\altaffilmark{3}, 
W.~B.~Atwood\altaffilmark{4}, 
M.~Axelsson\altaffilmark{5,6}, 
L.~Baldini\altaffilmark{7}, 
J.~Ballet\altaffilmark{8}, 
G.~Barbiellini\altaffilmark{9,10}, 
D.~Bastieri\altaffilmark{11,12}, 
B.~M.~Baughman\altaffilmark{13}
% ...
}
\altaffiltext{1}{National Research Council Research Associate}
\altaffiltext{2}{Space Science Division, Naval Research Laboratory, Washington, DC 20375}
\altaffiltext{3}{W. W. Hansen Experimental Physics Laboratory, Kavli Institute for Particle Astrophysics and Cosmology, Department of Physics and SLAC National Accelerator Laboratory, Stanford University, Stanford, CA 94305}
\altaffiltext{4}{Santa Cruz Institute for Particle Physics, Department of Physics and Department of Astronomy and Astrophysics, University of California at Santa Cruz, Santa Cruz, CA 95064}
\altaffiltext{5}{The Oskar Klein Centre for Cosmo Particle Physics, AlbaNova, SE-106 91 Stockholm, Sweden}
\altaffiltext{6}{Department of Astronomy, Stockholm University, SE-106 91 Stockholm, Sweden}
\altaffiltext{7}{Istituto Nazionale di Fisica Nucleare, Sezione di Pisa, I-56127 Pisa, Italy}
\altaffiltext{8}{Laboratoire AIM, CEA-IRFU/CNRS/Universit\'e Paris Diderot, Service d'Astrophysique, CEA Saclay, 91191 Gif sur Yvette, France}
\altaffiltext{9}{Istituto Nazionale di Fisica Nucleare, Sezione di Trieste, I-34127 Trieste, Italy}
\altaffiltext{10}{Dipartimento di Fisica, Universit\`a di Trieste, I-34127 Trieste, Italy}
\altaffiltext{11}{Istituto Nazionale di Fisica Nucleare, Sezione di Padova, I-35131 Padova, Italy}
\altaffiltext{12}{Dipartimento di Fisica ``G. Galilei", Universit\`a di Padova, I-35131 Padova, Italy}
\altaffiltext{13}{Department of Physics, Center for Cosmology and Astro-Particle Physics, The Ohio State University, Columbus, OH 43210}
% ...

\begin{abstract}
We report the discovery of $\gamma$-ray pulsations ($\ge 0.1$ GeV) from the young radio and X-ray pulsar PSR~J0205$+$6449 located in the Galactic supernova remnant 3C~58. Data in the $\gamma$-ray band were acquired by the Large Area Telescope aboard the \textit{Fermi Gamma-ray Space Telescope} (formerly GLAST), while the radio rotational ephemeris used to fold $\gamma$-rays was obtained using both the Green Bank Telescope and the Lovell telescope at Jodrell Bank. The light curve consists of two peaks separated by $0.49 \pm 0.01 \pm 0.01$ cycles which are aligned with the X-ray peaks. The first $\gamma$-ray peak trails the radio pulse by $0.08 \pm 0.01 \pm 0.01$, while its amplitude decreases with increasing energy as for the other $\gamma$-ray pulsars. Spectral analysis of the pulsed $\gamma$-ray emission suggests a simple power law of index $-2.1 \pm 0.1 \pm 0.2$ with an exponential cut-off at $3.0^{+1.1}_{-0.7} \pm 0.4$ GeV. The first uncertainty is statistical and the second is systematic. The integral $\gamma$-ray photon flux above 0.1 GeV is $(13.7 \pm 1.4 \pm 3.0)\times 10^{-8}$\,cm$^{-2}$\,s$^{-1}$, which implies for a distance of 3.2\,kpc and assuming a broad fan-like beam a luminosity of $8.3 \times 10^{34}$\,ergs\,s$^{-1}$ and an efficiency $\eta$ of 0.3\%. Finally, we report a 95\% upper limit on the flux of 1.7 $\times$ 10$^{-8}$\,cm$^{-2}$\,s$^{-1}$ for off-pulse emission from the object.
\end{abstract}


\section{INTRODUCTION}

The radio source 3C~58 was recognized early to be a supernova remnant (SNR G130.7+3.1; \citealt{Caswell70}), and subsequently classified as a pulsar wind nebula (PWN, also plerion) by \citep{Weiler78}. \citet{Becker82} identified an X-ray point source in the heart of 3C~58 as a likely pulsar, and subsequent studies yielded a distance of 3.2 kpc \citep{Rob93}. The pulsar J0205+6449 was finally discovered in {\em Chandra X-ray Observatory\/} data, with a period of 65.7 ms, while {\em Rossi X-ray Timing Explorer\/} archives allowed a measurement of the spindown rate of $\dot P = 1.93\times10^{-13}$ \citep{Murray02}. Both telescopes observed an X-ray profile with two narrow peaks separated by 0.5 in phase. This was followed by the detection of weak radio pulsations with a pulse averaged flux density of $\sim 45\,\mu$Jy at 1.4\,GHz and a sharp pulse of width 2\,ms \citep{Camilo02}. The pulsar has a very high spin-down luminosity of $2.7 \times 10^{37}$\,ergs\,s$^{-1}$ (the third most energetic of the known Galactic pulsars), a surface magnetic field strength of $3.6 \times 10^{12}$\,G, and a characteristic age of 5400 years. It also exhibits a high level of timing noise \citep{Ransom04}, and at least two glitches have occurred since its discovery \citep{Livingstone08}. Recently, a study by \citet{Livingstone09} presented the first measurement of the phase offset between the radio and X-ray pulse, showing that the first X-ray peak lags the radio pulse by $\phi=0.10\pm0.01$. 

The 3C~58/J0205+6449 system coincides positionally with the 828-year-old historical supernova SN~1181 according to \citet{Stephenson71} and \citet{Stephenson02}. However, recent investigations of models for the PWN \citep{Chevalier05} and the velocities of both the radio expansion \citep{Bie06} and optical knots \citep{Fesen08} imply an age for 3C~58 of several thousand years, closer to the characteristic age of the pulsar than of SN~1181. 3C 58 is similar to the apparently comparably aged Crab nebula (remnant of SN 1054) both having a flat-spectrum radio nebula, non-thermal extended X-ray emission, and point-like X-ray emission due to a central pulsar. But the two objects differ significantly both in luminosity and in size. The radio nebula of 3C~58, although $\sim 2$ times larger, is less luminous than the Crab by an order of magnitude \citep{Ivanov04}, while its X-ray luminosity is $\sim 2000$ times smaller \citep{Torii00}. These disparities could be explained by a different age. 

\section{OBSERVATIONS}

\subsection{Radio Timing Observations}
As part of the space and ground based timing program supporting {\em Fermi} \citep{Smith08}, PSR~J0205+6449 is being observed at the NRAO Green Bank Telescope (GBT) and the Lovell telescope at Jodrell Bank. The GBT provides higher precision data, but observations are made more frequently at Jodrell Bank, and the rotational ephemeris used here to fold gamma-ray photons is based on times-of-arrival (TOAs) obtained from both telescopes between 2008 June 17 and 2009 March 9. There are 17 such TOAs from GBT mainly at a central frequency of 2.0\,GHz, with average uncertainty of 0.27\,ms, each based on a 1\,hr integration. The 51 Jodrell Bank TOAs, with average uncertainty of 0.37\,ms, are derived from observations at 1.4\,GHz lasting typically for 2\,hr. The timing of PSR~J0205+6449 is very noisy, and in order to describe its rotation well during the 9 month interval we use TEMPO\footnote{http://www.atnf.csiro.au/research/pulsar/tempo/} to fit to the rotation frequency and its first seven derivatives, with an rms of 0.4\,ms. The best determination of dispersion measure (DM = $140.7 \pm 0.3$\,pc\,cm$^{-3}$) remains that from \citet{Camilo02}, which we use to correct 2\,GHz arrival times to infinite frequency, with an uncertainty of 0.3\,ms, for comparison with the gamma-ray profile. Note that higher signal-to-noise ratio profiles of the pulsar are presented at 0.8 and 1.4 GHz in \citet{Camilo02}.

\subsection{Gamma-Ray Observations}

The Large Area Telescope (LAT) on \textit{Fermi}, launched on 2008 June 11, is a pair-conversion telescope consisting of 16 towers set into a $4\times4$ grid \citep{Atwood09}. Each tower consists of both a converter-tracker (direction measurement of the incident $\gamma$-rays) and a CsI(Tl) crystal calorimeter (energy measurement/shower development image). The array is surrounded by segmented plastic scintillator (charged-particle background identification) and connected to a programmable trigger and data acquisition system. The instrument is sensitive to photons from 0.02 to 300 GeV over a $\sim 2.4$ sr field of view resulting from the compact height/width ratio of the instrument. The LAT hardware design, event reconstruction algorithms, background selections and event quality selections determine the instrument
performance\footnote{http:$//$www$-$glast.slac.stanford.edu$/$software$/$IS$/$glast\_lat\_performance.htm}: 
a large effective area on axis ($\sim 0.8$ m$^{2}$); superior angular resolution ($\theta_{68} \sim 0.5^{\circ}$ at 1 GeV for events in the front section of the tracker); and an energy resolution better than 10\% between 0.1 and 10 GeV on axis. The software timing chain deriving from a GPS clock on the satellite and the phase-folding software have been shown to be accurate to better than a few $\mu$s \citep{Smith08}.

In this letter, the data collected for the timing analysis were obtained by the \textit{Fermi} LAT in two different observing modes, from 2008 July 15 to July 29 during the six-week calibration phase, and from 2008 August 3 to 2009 March 9 when LAT was operating in scanning mode under nominal configuration. For the spectral analysis, only data acquired from 2008 August 3 were used to avoid mixing the different configurations. We used the ``Diffuse'' class events having the highest probability of being photons. In addition, we excluded the events with zenith angles greater than $105{^\circ}$ due to the Earth's bright $\gamma$-ray albedo, and time periods where the Earth's limb came within 28$^{\circ}$ of the source.


\section{ANALYSIS AND RESULTS}

The events were analyzed using the standard software package \textit{Science Tools} (ST) for the \textit{Fermi} LAT data analysis\footnote{http:$//$fermi.gsfc.nasa.gov/ssc/data/analysis/scitools/overview.html} and the pulsar timing software TEMPO2 \citep{Hobbs06}. The timing parameters used in this work will be made available on the servers of the \textit{Fermi} Science Support Center\footnote{http:$//$fermi.gsfc.nasa.gov/ssc/data/access/lat/ephems/}.

\subsection{Pulsed Light Curve}

The pulsar is located in the Galactic plane where the diffuse gamma radiation is intense, and $5.3^{\circ}$ from the bright LAT source 0FGL~J0240.3+6113 coincident with the X-ray binary LSI+61$^{\circ}$303 \citep{Abdo09a}. For the timing analysis, a set of photons with energies over 0.1 GeV was selected within an energy-dependent cone of radius $\theta_{68} \leqslant$ $0.8 \times E_{\rm GeV}^{-0.75}$\,degrees, but with a maximum radius of $1.5^{\circ}$ 
with respect to the X-ray pulsar position ($l=130.719^{\circ}$, $b=3.085^{\circ}$). This choice takes into account the instrument performance and maximizes the signal-to-noise ratio over a broad energy range. This truncates the point spread function at low energies and decreases the number of background events \citep{Atwood09}. A total of 2922 $\gamma$-rays remain after these cuts. The arrival times of events were corrected to the Solar System Barycenter using the JPL DE405 Solar System ephemeris \citep{standish98}, and the events have been folded using the radio ephemeris from GBT and the Lovell telescope at Jodrell Bank. 

\begin{figure*}
\epsscale{1.0}
\plotone{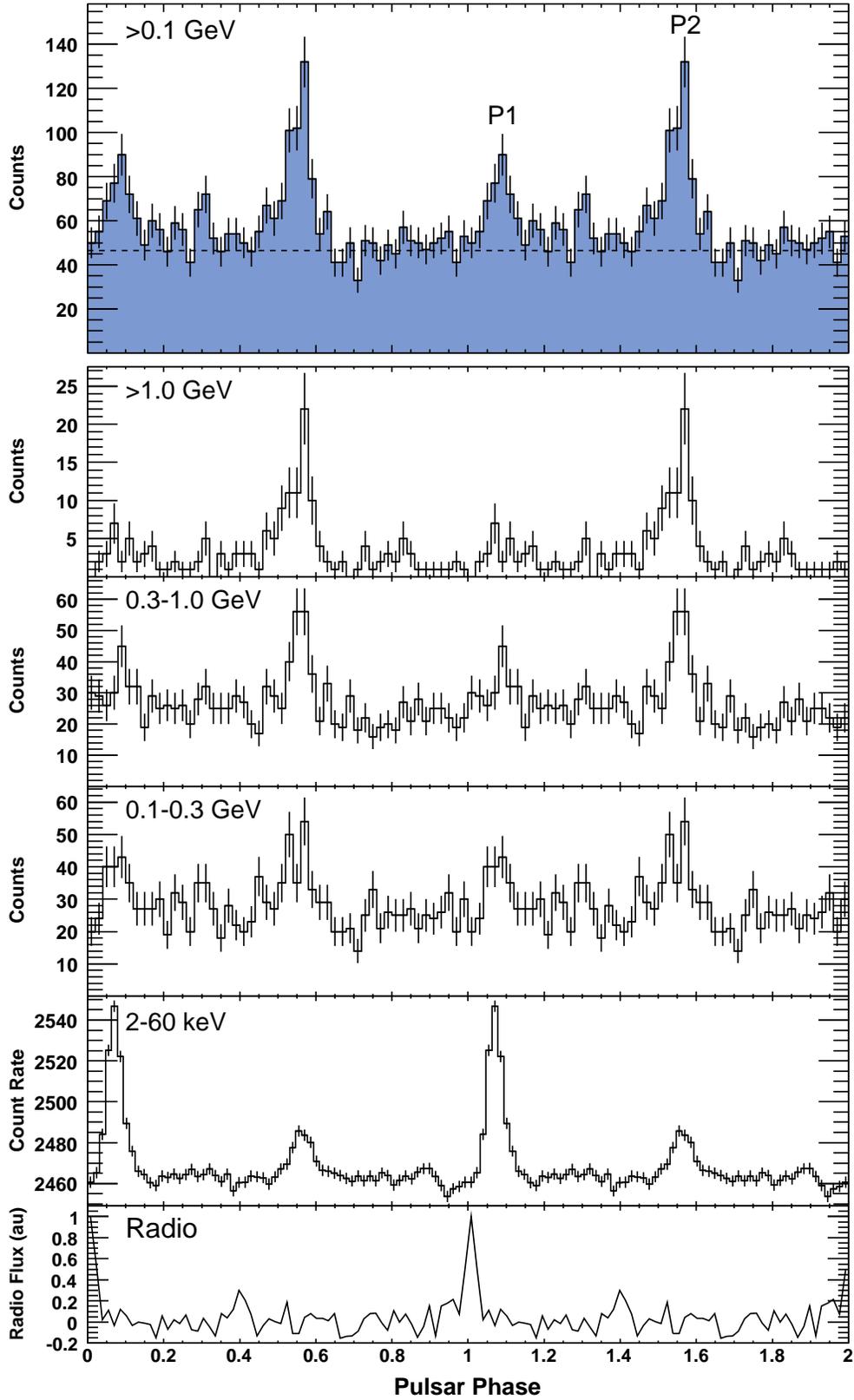}
\caption{
\textbf{Top panel:}~Phase-aligned histogram of PSR~J0205+6449 above 0.1 GeV and within an energy-dependent circle. Two rotations are plotted with 50 bins per period. The dashed line shows the background level, as estimated from a ring surrounding the pulsar during the off-pulse phase (46 counts/bin). 
\textbf{Three following panels:}~Energy dependent phase histograms for PSR~J0205+6449 in the three indicated energy ranges, each displayed with 50 bins per pulse period. 
\textbf{Second panel from bottom:}~Count rate in the energy band 2$-$60 keV from \textit{RXTE} data \citep{Livingstone09}.
\textbf{Bottom panel:}~Radio pulse profile based on 3.8 hours of GBT observations at a center frequency of 2\,GHz with 64 phase bins. \label{fig1} }
\end{figure*}

Figure~\ref{fig1} (top panel) shows the 50 bin $\gamma$-ray phase histogram on which we fit each peak. The first peak (P1) is offset from the radio pulse (bottom panel) by $0.08 \pm 0.01 \pm 0.01$ according to a Lorentzian fit with a full width at half maximum (FWHM) of $0.15 \pm 0.03$. The first phase uncertainty arises from the $\gamma$-ray fit, and the second is from the DM uncertainty in extrapolating the radio time-of-arrival to infinite frequency. The second peak (P2) is asymmetric, fit by a two-sided Lorentzian to take into account the different widths for the leading and trailing edges. The fit places the peak at $0.57 \pm 0.01 \pm 0.01$ with an FWHM of $0.13 \pm 0.04$. The two peaks are separated by $0.49 \pm 0.01 \pm 0.01$ in phase. We defined the `off pulse region' as the pulse minimum between 0.65 and 1.0 in phase. The dashed line in Figure~\ref{fig1} represents the background counts measured from a 2\ --\ 3$^{\circ}$ ring surrounding the pulsar during the pulse minimum. Evidence for an excess between P1 and P2 appears with a significance ${\rm (signal/\sqrt{background})}$ of $5.0\,\sigma$ between 0.14 and 0.46 in phase. However, the data cannot constrain the pulse phase dependence of this excess flux.

To examine the energy-dependent trend of the $\gamma$-ray pulse profile, 50 bin phase histograms are plotted in three energy intervals, 0.1\ --\ 0.3 GeV, 0.3\ --\ 1 GeV, and $\ge$1 GeV in Figure~\ref{fig1} (middle panels), which show an evolution in the shape. Between 0.1 and 1 GeV, we see distinctly two peaks and a possible bridge region in excess of the unpulsed emission, while above 1 GeV only P2 is significant. We also observe that the ratio P1/P2 (sum of counts) decreases with increasing energy, with a ratio of $0.63\pm0.06$ between 0.1\ --\ 0.3 GeV, $0.55\pm0.05$ between 0.3\ --\ 1.0 and $0.24 \pm 0.06$ above 1 GeV. The Vela \citep{Abdo09b}, Crab, Geminga and B1951+32 pulsars show similar behavior \citep{Thompson01}. This general trend suggests a spectral energy dependence of the $\gamma$-ray light curve. Finally, we note that the highest energy photon is in P2 with an energy of 8.6 GeV.

Figure~\ref{fig1} (bottom panels) also shows the 2\ --\ 60 keV X-ray phase histogram measured by \textit{RXTE} \citep{Livingstone09} as well as the 2 GHz radio profile from GBT (the single peak defines phase $\phi = 0$). \citet{Livingstone09} find that the radio pulse leads the X-ray pulse by $\phi = 0.10 \pm 0.01$ and note a separation between the two narrow X-ray peaks of $\Delta\phi = 0.5$. The good alignment between the X-ray and $\gamma$-ray profiles suggests a common origin between the two components. This feature is also observed in the LAT data for the Vela pulsar where the more intense X-ray pulse is aligned with the first $\gamma$-ray pulse \citep{Abdo09b}.


\subsection{Spectra and phase-averaged flux}

To study the phase-averaged spectrum of J0205$+$6449, a maximum likelihood spectral analysis\footnote{http://fermi.gsfc.nasa.gov/ssc/data/analysis/documentation/} \citep{Mattox96}, implemented in the ST as the \textit{gtlike} task, was performed using a $20{^\circ}$ $\gamma$-ray map centered on the pulsar position between 0.1  and 200 GeV. The systematic errors on the effective area are currently estimated as $\le$ 5\% near 1 GeV, 10\% below 0.1 GeV and 20\% above 10 GeV. The diffuse emission from the Galactic plane was modeled using maps based on the GALPROP model (gll\_iem\_v01) \citep{Strong04a,Strong04b}. The extragalactic radiation as well as the instrumental backgrounds were modeled by an isotropic component with a power-law spectral shape. Nearby LAT sources were included in the analysis. For PSR~J0205+6449, we modeled the shape of the spectrum with a power law with an exponential cut-off.

We first determined the diffuse background components by selecting off-pulse data. Then, we fitted the on-pulse data using the diffuse background spectrum thus obtained in order to improve the signal-to-background ratio of the pulsar and the nearby sources. The best fit result is described by:
\begin{eqnarray}
\frac{dF}{dE} = N_{0}\ E^{-\Gamma}\ e^{-E/E_{c}}
\label{eqn_diff_spec}
\ {\rm cm^{-2} s^{-1} GeV^{-1}}
\end{eqnarray}
with E in GeV, the term $N_{0} = (1.4 \pm 0.1 \pm 0.1) \times 10^{-8}$\,cm$^{-2}$\,s$^{-1}$\,GeV$^{-1}$, the spectral index $\Gamma = 2.1 \pm 0.1 \pm 0.2$, and the cut-off energy $E_{c} = 3.0^{+1.1}_{-0.7} \pm 0.4$ GeV. The errors are the statistical and propagated systematic uncertainties, respectively. We obtain from this fit over the range 0.1--200 GeV an integral photon flux of $(13.7 \pm 1.4 \pm 3.0) \times 10^{-8}$\,cm$^{-2}$\,s$^{-1}$ and integral energy flux of $F_{E,obs} = (6.7 \pm 0.5 \pm 1.0) \times 10^{-11}$\,ergs\,cm$^{-2}$\,s$^{-1}$. To check the assumption of a cut-off energy in the spectrum, we have also fit the same dataset with a simple power-law of the form $dF/dE = N_{0}(E/1\rm GeV)^{-\Gamma}$. The spectral model using an exponential cut-off is better constrained with a difference between the log likelihoods of $\sim 4.5\,\sigma$, disfavoring the power-law hypothesis. The source, having a statistical significance of 9.5 $\sigma$ ($<$ 10 $\sigma$) from 0.2 to 100 GeV for the first three months of the sky survey in the science phase of the mission, does not appear in the \textit{Fermi} LAT bright $\gamma$-ray source list \citep{Abdo09a}.

As a first search for unpulsed emission from the nebula 3C~58, we fitted a point-source to the off-pulse data at the radio pulsar position in the energy band 0.2\ --\ 200 GeV. No signal was observed from the PWN. Lastly, after scaling to the full pulse phase, we derived a 95\% CL upper limit on the flux of $1.7\times 10^{-8}$\,cm$^{-2}$\,s$^{-1}$.


\section{DISCUSSION}


\subsection{Light curve}

Multi-wavelength light curves are important to locate the pulse emission in the open field line region and hence to understand the mechanisms of particle acceleration. The $\gamma$-ray profile for J0205+6449, covering a wide range in phase, is reminiscent of the Vela light curve \citep{Abdo09b}, including the $\gamma$-ray peak alignment with the X-ray pulses. The $\gamma$-ray delay of 0.08 cycles and the peak separation of $\sim 0.5$ is becoming a consistent pattern, as the first $\gamma$-ray peaks for Vela, B1951+32, and J2021+3651 \citep{Halpern08,Abdo09c} lag the radio pulses by 0.13, 0.16, and 0.17 respectively, and the separation of the $\gamma$-ray peaks is 0.4\ --\ 0.5. This fits the predictions of the outer magnetospheric models quite well, whether they be the traditional outer gap model (OG) \citep{Romani95} or the two pole caustic gap model (TPC) \citep{Dyks03}.

\subsection{Beaming and Luminosity}

To know the $\gamma$-ray efficiency, we need to determine the total luminosity $L_{\gamma}= 4\pi f_{\Omega}$($\alpha$,$\zeta_E$) $F_{E,obs} D^{2}$, where $D$ is the distance to the pulsar, $ F_{E,obs}$ is the observed phase-averaged energy flux for the Earth line of sight (at angle $\zeta_E$ to the rotation axis), and $f_{\Omega}$($\alpha$,$\zeta_E$) is a correction factor that takes into account the beaming geometry, given by \citep{Watters08}:
\begin{equation}\label{eq:fbeam}
f_{\Omega}(\alpha,\zeta_{E}) = \frac{\int F_{\gamma}(\alpha;\zeta,\phi)sin(\zeta)d\zeta d\phi}{2 \int F_{\gamma}(\alpha;\zeta_{E},\phi) d\phi}
\end{equation}
where $F_{\gamma}(\alpha;\zeta,\phi)$ is the radiated flux as a function of the viewing angle ($\zeta$) and the pulsar phase ($\phi)$, while $f_{\Omega}$ is the ratio between the overall $\gamma$-ray emission over the full sky and the expected phase-averaged flux for the light curve seen from the Earth. In the case of a polar cap model \citep{Harding05} where charged particles are accelerated in charge-depleted zones near the poles of the pulsar, $f_{\Omega} \ll 1$, while for the outer magnetosphere models the emission is radiated with $f_{\Omega} \gtrsim 1$. 

An important uncertainty when evaluating the $\gamma$-ray efficiency arises from the determination of the distance. Observations of neutral hydrogen (H~I) absorption by \citet{Rob93} yielded an estimate for the radial velocity of 3C~58. They convert this to a kinematical distance of 3.2 kpc assuming a flat Galactic rotation curve as per \citet{Fich89}, who quotes distance uncertainties of order 25\%, making this result consistent with the 2.6 kpc previously reported by \citet{Green82}. The distance derived from the DM is 4.5 kpc, according to the NE 2001 model \citep{Cordes02}, with an uncertainty that can exceed 50\%, depending on the viewing direction, consistent with the kinematical distance.

We estimate $L_{\gamma}= 8.3 \times 10^{34}\ (D/3.2\,{\rm kpc})^2\ f_{\Omega}$\,ergs\,s$^{-1}$, and deduce, for a neutron star moment of inertia of $10^{45}$\,g\,cm$^{2}$, a $\gamma$-ray efficiency $\eta = L_{\gamma}/\dot{E} = 0.003\ f_{\Omega}\ (D/3.2\,{\rm kpc})^2$. Assuming the outer magnetospheric models for PSR~J0205+6449, we can deduce the parameters $\alpha$ and $\zeta$ from the $\gamma$-ray peak separation and using the $\gamma$-ray light curve ``Atlas'' of \citet{Watters08}. For the OG model, we estimate $f_{\Omega}\approx 0.9$--$1.0$ with $\alpha \backsim 60^{\circ}$--90$^{\circ}$ and $\zeta \backsim 80^{\circ}$--85$^{\circ}$, while for the TPC model, $f_{\Omega} \approx 0.95$--1.25 with both $\alpha \backsim 50^{\circ}$--$90^{\circ}$, $\zeta \backsim 85^{\circ}$--90$^{\circ}$ and $\alpha \backsim 85^{\circ}$--$90^{\circ}$, $\zeta \backsim 45^{\circ}$--90$^{\circ}$. Examination of the geometry of the PWN 3C 58 using {\em Chandra\/} \citep{Ng04,Ng08} yielded a viewing angle of $\zeta = 91.6 \pm 0.2 \pm 2.5$ $^\circ$ (inner torus) based on the tilt angle of the torus to the plane of the sky. This value is in agreement with the results from the OG model and more consistent with the first estimated range for the TPC model. Adopting an average value for $f_{\Omega} = 1$ according to the OG model and a distance of 3.2~kpc, we evaluate a $\gamma$-ray efficiency $\eta = 0.3\%$ for converting its rotational energy loss into $\gamma$-rays. 

Studies of rotation powered pulsars suggest that the efficiency $\eta$ for converting its rotational energy loss into $\gamma$-rays increases as the open field line voltage $V \backsimeq 4 \times 10^{20} P^{-3/2}\dot{P}^{1/2}$ volts decreases and is proportional to $\dot{E}^{-1/2}$ \citep{Arons96}. V is also proportional to the open field current \citep{Harding81} and to the characteristic age of the pulsar \citep{Buccheri78}. We note that with an open field line voltage of $1.0 \times 10^{16}$\,V, that PSR~J0205+6449 closely follows the approximate $\eta \varpropto \dot{E}^{-1/2} \varpropto 1/V$ relations, confirming the trend.  

\citet{Thompson99} point out that for the $\gamma$-ray EGRET pulsars the broadband energy spectra for the young pulsars like Crab and B1509$-$58 peaks in the X-ray band, while the older $\gamma$-ray pulsars have their maximum luminosity in the high-energy regime. Considering a X-ray luminosity between 0.5 and 8 keV of $1.51 \times 10^{33}$\,ergs\,s$^{-1}$ \citep{Kargaltsev08} and a gamma-ray luminosity of $8.3 \times 10^{34}$\,ergs\,s$^{-1}$ with $f_{\Omega} = 1$, PSR~J0205+6449 seems to have an $L_{X}/L{\gamma}$ ratio closer to the middle-aged Vela pulsar than the young Crab and B1509$-$58 pulsars, and hence suggests that the association of 3C~58 and historical supernova SN~1181 is probably incorrect.

\section{SUMMARY}

Using a rotational ephemeris derived from radio observations with the GBT and at Jodrell Bank, and $\gamma$-ray data from the \textit{Fermi} LAT, we have discovered in the Galactic SNR 3C~58 $\gamma$-ray pulsations from PSR~J0205+6449, the third most energetic of the known Galactic pulsars. This source has no EGRET counterpart and presents a new opportunity to study the pulsar/SNR association at high energies, demonstrating the good performance of the instrument.

1.~The $\gamma$-ray profile for J0205+6449 is similar to the majority of known $\gamma$-ray pulsar light curves, in particular to the Vela pulsar. It consists of two peaks separated by $0.49 \pm 0.01 \pm 0.01$ in phase. The radio pulse leads the $\gamma$-ray pulse by $\phi$ = $0.08 \pm 0.01 \pm 0.01$ and the X-ray peaks are aligned with the $\gamma$-ray peaks.

2.~We report a 95\% CL upper limit on the flux of 1.7 $\times$ 10$^{-8}$\,cm$^{-2}$\,s$^{-1}$ for a possible unpulsed emission from the nebula 3C~58.

3.~The $\gamma$-ray energy spectrum above 0.1 GeV can be described by a simple power law with photon index of $2.1 \pm 0.1 \pm 0.2$ and an energy cut-off of $3.0^{+1.1}_{-0.7} \pm 0.4$ GeV.

4.~Adopting a distance of 3.2~kpc and assuming a broad fan-like beam, we report an efficiency of $\sim$0.3\% for the conversion of spin-down energy into $\gamma$-ray emission.

\acknowledgements
The $Fermi$ LAT Collaboration acknowledges support from a number of agencies and institutes for both development and the operation of the LAT as well as scientific data analysis. These include NASA and DOE in the United States, CEA/Irfu and IN2P3/CNRS in France, ASI and INFN in Italy, MEXT, KEK, and JAXA in Japan, and the K.~A.~Wallenberg Foundation, the Swedish Research Council and the National Space Board in Sweden. Additional support from INAF in Italy for science analysis during the operations phase is also gratefully acknowledged.

Additional support for science analysis during the operations phase from the following agencies is also gratefully acknowledged:  the Istituto Nazionale di Astrofisica in Italy and the K.~A. Wallenberg Foundation in Sweden for providing a grant in support of a Royal Swedish Academy of Sciences Research fellowship for JC.

The Green Bank Telescope is operated by the National Radio Astronomy Observatory, a facility of the National Science Foundation operated under cooperative agreement by Associated Universities, Inc.

The Lovell Telescope is owned and operated by the University of Manchester as part of the Jodrell Bank Centre for Astrophysics with support from the Science and Technology Facilities Council of the United Kingdom.

\end{document}